\documentclass[conference]{IEEEtran}
\IEEEoverridecommandlockouts
\usepackage{cite}
\usepackage{amsmath,amssymb,amsfonts}
\usepackage[ruled, lined, vlined]{algorithm2e}
\usepackage{algorithmic}
\usepackage{graphicx}
\usepackage{textcomp}
\usepackage{xcolor}
\usepackage{booktabs}
\def\BibTeX{{\rm B\kern-.05em{\sc i\kern-.025em b}\kern-.08em
    T\kern-.1667em\lower.7ex\hbox{E}\kern-.125emX}}

\begin{document}

\title{Deep Reinforcement Learning based Recommendation with Explicit User-Item Interactions Modeling
}



\author{\IEEEauthorblockN{Feng Liu\IEEEauthorrefmark{1}, Ruiming Tang\IEEEauthorrefmark{2}, Xutao Li\IEEEauthorrefmark{1}, Weinan Zhang\IEEEauthorrefmark{3}\ Yunming Ye\IEEEauthorrefmark{1}, Haokun Chen\IEEEauthorrefmark{3}, Huifeng Guo\IEEEauthorrefmark{2} and Yuzhou Zhang\IEEEauthorrefmark{2}}
\IEEEauthorblockA{\IEEEauthorrefmark{1}Shenzhen Key Laboratory of Internet Information Collaboration\\
Shenzhen Graduate School, Harbin Institute of Technology, Shenzhen, 518055, China\\
Email: fengliu@stu.hit.edu.cn, lixutao@hit.edu.cn, yeyunming@hit.edu.cn}
\IEEEauthorblockA{\IEEEauthorrefmark{2}Noah's Ark Lab, Huawei, China\\
Email: tangruiming, huifeng.guo, zhangyuzhou3@huawei.com}
\IEEEauthorblockA{\IEEEauthorrefmark{3}Shanghai Jiao Tong University, Shanghai, China\\
Email: wnzhang@sjtu.edu.cn, chenhaokun@sjtu.edu.cn}
}

\maketitle

\begin{abstract}
Recommendation is crucial in both academia and industry, and various techniques are proposed such as content-based collaborative filtering, matrix factorization, logistic regression, factorization machines, neural networks and multi-armed bandits. However, most of the previous studies suffer from two limitations: (1) considering the recommendation as a static procedure and ignoring the dynamic interactive nature between users and the recommender systems; (2) focusing on the immediate feedback of recommended items and neglecting the long-term rewards. To address the two limitations, in this paper we propose a novel recommendation framework based on deep reinforcement learning, called DRR. The DRR framework treats recommendation as a sequential decision making procedure and adopts an ``Actor-Critic'' reinforcement learning scheme to model the interactions between the users and recommender systems, which can consider both the dynamic adaptation and long-term rewards. Further more, a state representation module is incorporated into DRR, which can explicitly capture the interactions between items and users. Three instantiation structures are developed. Extensive experiments on four real-world datasets are conducted under both the offline and online evaluation settings. The experimental results demonstrate the proposed DRR method indeed outperforms the state-of-the-art competitors.

\end{abstract}

\begin{IEEEkeywords}
Recommendation, Deep Reinforcement Learning, User-Item Interactions
\end{IEEEkeywords}

\section{Introduction}

Thanks to the increasing online services, such as online shopping, online news and online social networks, it becomes quite convenient to acquire items (goods, books, videos, news, etc.) via Internet or mobile devices. Albeit the great convenience, the overwhelming number of items in the systems also pose a significant challenge for users, to find the items that match their interests. Recommendation is a widely used solution and various families of techniques have been proposed, such as content-based collaborative filtering~\cite{contentCF}, matrix factorization based methods~\cite{itemBased,MFTeches,AmazonCF,unifyCF}, logistic regression, factorization machines and its variants~\cite{ftrl,fm,ffm}, deep learning models~\cite{fnn,pnn,deepfm,wide&deep} and multi-armed bandits~\cite{thompsonSampling,banditForNews,factorizationBandit,onlineContextMAB,interactiveCF}. However, such mentioned studies suffer from two serious limitations. 



\begin{figure}[ht]
\centering
\begin{minipage}[b]{0.5\textwidth}\centering
\includegraphics[width=0.45\textwidth]{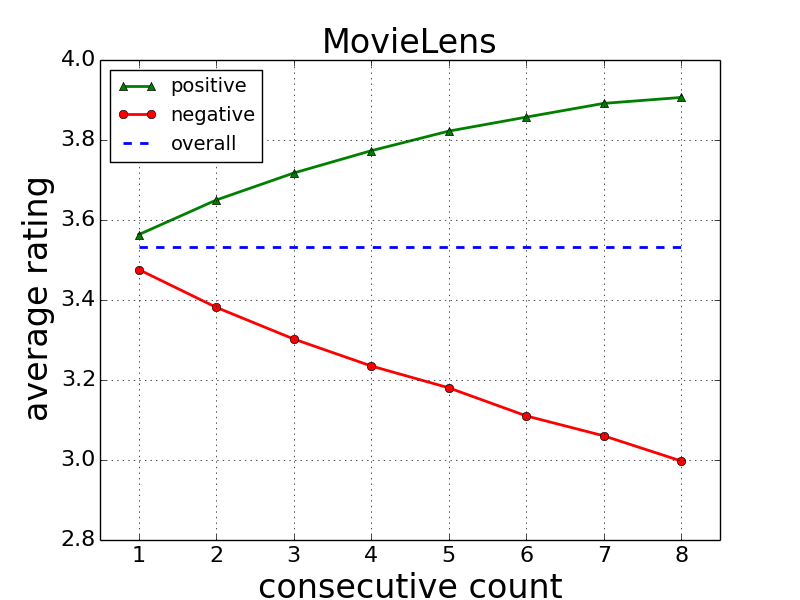}
\includegraphics[width=0.45\textwidth]{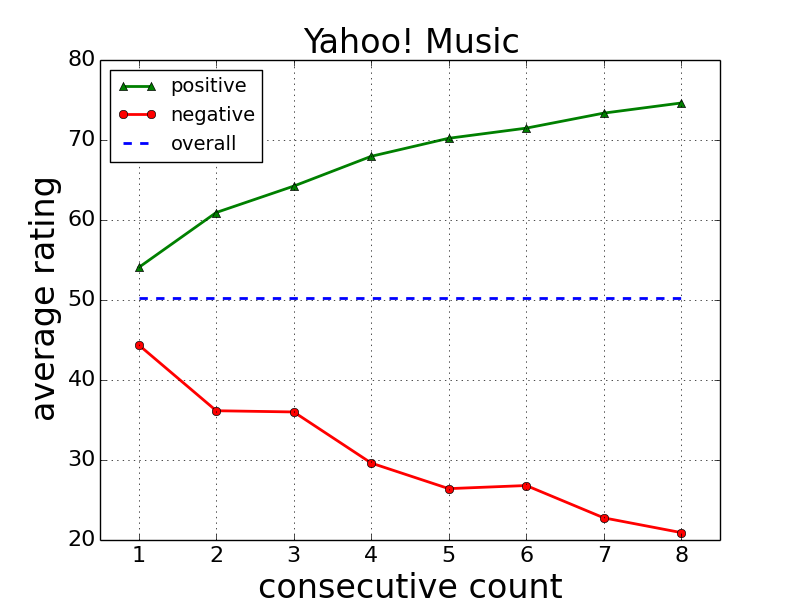}
\end{minipage}
\caption{Analysis on sequential patterns on user's behavior in MovieLens and Yahoo!Music datasets}\label{fig:sequential-pattern}
\end{figure}

\emph{Firstly}, most of them consider the recommendation procedure as a static process, i.e., they assume the user's underlying preference keeps unchanged. However, it is very common that a user's preference is dynamic w.r.t. time, i.e., a user's preference on previous items will affect her choice on the next items. Hence, it would be more reasonable to model the recommendation as a sequential decision making process. We will show some evidence observed in publicly available datasets (MovieLens and Yahoo! Music) to support our opinion. In the two datasets, the sequential behaviors of users are recorded and we are interested in what would happen if a user consecutively receives satisfied or unsatisfied recommendations. Though the datasets do not record any recommendation procedure, we can simulate this according to the users' ratings, namely, consecutive rating ``positive'' (``negative'') simulates that a user consecutively receives satisfied (unsatisfied) recommendations. As presented in Figure \ref{fig:sequential-pattern}, we observe that a user tends to gives a higher (lower) rating if she has consecutively received more satisfied (unsatisfied) items, as shown by the green (red) line, where the blue dot line denotes the average rating for reference. This suggests that a user will be more pleasant (unpleasant) if she consecutively receives more satisfied (unsatisfied) recommendations and therefore she tends to give a higher (lower) rating to the current recommendation. Hence, the user's dynamic preference suggests that a good recommendation should be modeled as a sequential decision making process.

\emph{Secondly}, the aforementioned studies are trained by maximizing the immediate rewards of recommendations, which merely concentrates on whether the recommended items are clicked or consumed, but ignores the long-term contributions that the items can make. However, the items with small immediate rewards but large long-term benefits are also crucial~\cite{MDPRec}. We take an example in News recommendation~\cite{NewsRL} to explain this. As a user requests for news to read, two possible pieces of news may lead to the same immediate reward, i.e., the user will click and read the two pieces of news with equal probability, where one is about a thunderstorm alert and the other is about a basketball player Kobe Bryant. In this example, after reading the news about thunderstorm, the user probably is not willing to read news about this issue anymore; while on the other hand, the user will possibly read more about NBA or basketball after reading the news about Kobe. The fact suggests that recommending the news about Kobe will introduce more long-term rewards. Hence, when recommending items to users, both the immediate and long-term rewards should be taken into consideration.

Recently, Reinforcement Learning (RL)~\cite{sutton1998reinforcement}, which has shown great potential in various challenging scenarios that require both dynamic modeling and long term planning, such as game playing~\cite{davidHuman,davidGo}, real-time ads bidding~\cite{RTB-1,RTB-2}, neural network structure searching~\cite{NNStructure-1,NNStructure-2}, is introduced in recommender systems~\cite{MDPRec,qlearnigRec,JDListwise,JDNegFeedback,NewsRL,SearchRL-XU-Long,SearchRL-XU-Short,SearchRL-Ali,DDPG-LargeAcionSpace}.


In the early stage, \emph{model-based RL} techniques are proposed to model recommendation procedure, such as POMDP~\cite{MDPRec} and Q-learning~\cite{qlearnigRec}. However, these methods are inapplicable to complicated recommendation scenarios when the number of candidate items is large, because a time-consuming dynamic programming step is required to update the model. Later, \emph{model-free RL} techniques are utilized in recommender systems, from both academia and industry. Such techniques can be divided into two categories: value-based~\cite{JDNegFeedback,NewsRL} and policy-based~\cite{JDListwise,SearchRL-Ali,DDPG-LargeAcionSpace}. Value-based approaches compute Q-values of all available actions for a given state and the one with the maximum Q-value is selected as the best action. Due to the evaluation on overall actions, the approaches may become very inefficient if the action space is too large. As for the policy-based approaches, this type of studies generate a continuous parameter vector as the representation of an action~\cite{JDListwise,SearchRL-Ali,DDPG-LargeAcionSpace}, which can be utilized in generating the recommendation and updating the Q-value evaluator. Thanks to the continuous representations, the inefficiency drawbacks can be overcome. However, these studies~\cite{JDListwise,SearchRL-Ali,DDPG-LargeAcionSpace} still have one common limitation: the user state is learnt via a conventional fully connected neural network, which does not explicitly and carefully model the interactions between users and items.



In this paper, to break the limitations stated above, we propose a \underline{d}eep \underline{r}einforcement learning based \underline{r}ecommendation framework with explicit user-item interactions modeling (DRR). The ``Actor-Critic'' type framework DRR is incorporated with a state representation module, which explicitly models the complex dynamic user-item interactions to pursuit better recommendation performance. Specifically, the embeddings of users and items from the historical interactions are fed into a carefully designed multi-layer network, which explicitly models the interactions between users and items, to produce a continuous state representation of the user in terms of her underlying sequential behaviors. This network is named as the state representation module, which plays two important roles in our framework. On the one hand, it is utilized to generate an ranking action to calculate the recommendation scores for ranking. On the other hand, the state representation together with the generated action is the input of the Critic network, which aims to estimate the Q-value, i.e., the quality of the action in the current state. Based on the evaluation, the Actor (policy) network can be updated. We note that both the Actor and Critic networks are carefully designed by modeling the interactions between users and items explicitly. Extensive experiments on four real-world datasets demonstrate that the proposed method yields superior performance than the state-of-the-art methods. The main contributions of this paper can be summarized as follows:

\begin{itemize}
\item We propose a deep reinforcement learning based recommendation framework DRR. Unlike the conventional studies, DRR adopts an ``Actor-Critic'' structure and treats the recommendation as a sequential decision making process, which takes both the immediate and long-term rewards into consideration.

\item Under the DRR framework, three different network structures are proposed, which can explicitly model the interactions between users and items.
\item Extensive experiments are carried out on four real-world datasets, and the results demonstrate the proposed methods indeed outperforms the state-of-the-art competitors.
\end{itemize}


The rest of this paper is organized as follows. Related work and background are presented in Section II. The preliminary knowledge is presented in Section III. The proposed methods are introduced in Section IV. Experimental details and results are discussed in Section V. Finally, we conclude this paper and discuss some future work in Section VI.

\section{Related Work}

\subsection{Non-RL based Recommendation Techniques}

Various kinds of recommendation techniques are proposed in the past a few decades to improve the performance of recommender systems, including content-based filtering~\cite{contentCF}, matrix factorization based methods~\cite{itemBased,MFTeches,AmazonCF,unifyCF}, logistic regression, factorization machines and its variants~\cite{ftrl,fm,ffm}, and until recently deep learning models~\cite{fnn,pnn,deepfm,wide&deep}.

At the beginning of this century, content-based filtering~\cite{contentCF} is proposed to recommend items by considering the content similarity between items. Later, collaborative filtering (CF) is put forward and extensively studied. The rationale behind CF is that the users with similar behaviors tend to prefer the same items, and the items consumed by similar users tend to have the same rating. However, conventional CF based methods tend to suffer from the data scarcity, because the similarity calculated from sparse data can be very unreliable. Matrix factorization (MF), as an advanced CF technique, plays an important role in recommender systems. MF models~\cite{itemBased,MFTeches,AmazonCF,unifyCF} characterize both items and users by vectors in the same space, which are inferred from the observed user-item interactions. Regarding the recommendation as a binary classification problem, logistic regression and its variants~\cite{ftrl} are also utilized in recommender systems. However, logistic regression based models are hard to generalize to the feature interactions that never or rarely appear in the training data. Factorization machines~\cite{fm} model pairwise feature interactions as inner product of latent vectors between features and show promising results. As an extension to FM, Field-aware FM (FFM~\cite{ffm}) enables each feature to have multiple latent vectors to interact with different fields. Recently, deep learning models~\cite{fnn,pnn,deepfm,wide&deep} are applied to model the complicated feature interactions for recommendation.


As a distinguished direction, contextual multi-armed bandits are also utilized to model the interactive nature of recommender systems~\cite{thompsonSampling,banditForNews,factorizationBandit,onlineContextMAB,interactiveCF}. Li et al. apply Thompson Sampling (TS) and Upper Confident Bound (UCB) to balance the trade-off between exploration and exploitation in~\cite{thompsonSampling} and~\cite{banditForNews}, respectively. The authors of~\cite{onlineContextMAB} propose a dynamic context drift model to address the time varying problem. To integrate the latent vectors of items and users with some exploration, the authors of~\cite{factorizationBandit,interactiveCF} combine matrix factorization with multi-armed bandits.

However, all these methods suffer from two limitations. First, they consider the recommendation procedure as a static process, i.e., they assume the underlying user's preference keeps static and they aim to learn the user's preference as precise as possible. Second, they are learned to maximize the immediate rewards of recommendations, but ignore the long-term benefits that the recommendations can make.

\subsection{RL based Recommendation Techniques}

As model-based RL techniques~\cite{MDPRec,qlearnigRec} are inapplicable in recommendation scenario due to their high time complexity, most researchers turn to model-free RL techniques. The model-free RL techniques can be divide into two categories: policy-based and value-based.

Policy-based approaches~\cite{JDListwise,SearchRL-Ali,DDPG-LargeAcionSpace} aim to generate a policy, of which the input is a state, and the output is an action. These works apply deterministic policies, which generates an action directly. Dulac-Arnold et al.~\cite{DDPG-LargeAcionSpace} resolves the large action space problem by modeling the state in a continuous item embedding space and selecting the items via a neighborhood method. However, as the underlying algorithm is essentially a continuous-action algorithm, its performance may be cursed by the gap between the continuous and discrete action spaces. In~\cite{JDListwise,SearchRL-Ali}, the policy network outputs a continuous action representation, and the recommendation is generated by ranking the items with their scores, which are computed by a pre-defined function with the action representation and the item embeddings as input. However, one common limitation of the studies is that they do not carefully learn the state representation.

For value-based approaches~\cite{JDNegFeedback,NewsRL}, the action with maximum Q-value over all the possible actions is selected as the best action. Zhao et al.~\cite{JDNegFeedback} take both user's positive feedback and negative feedback into consideration when modeling user state. Dueling Q-network is utilized in~\cite{NewsRL}, to model Q-value of a state-action pair. Moreover, a minor update with exploration by dueling bandit gradient descent is proposed. However, such value-based approaches need to evaluate the Q-values of all the actions under a specific state, which is very inefficient when the number of actions is large.

To make RL based recommendation techniques suitable for large-scale scenario, in this paper, we propose the DRR framework which carefully and explicitly model the interactions between users and items to learn the state representation.

\section{Preliminaries}

The essential underlying model of reinforcement learning is Markov Decision Process (MDP). An MDP is defined as $(\mathcal{S},\mathcal{A},\mathcal{P},\mathcal{R},\gamma)$. $\mathcal{S}$ is the state space and $\mathcal{A}$ is the action space. $\mathcal{P}:\mathcal{S}\times\mathcal{A}\times\mathcal{S}\mapsto [0,1]$ is the state transition function. $\mathcal{R}:\mathcal{S}\times\mathcal{A}\times\mathcal{S}\mapsto \mathbb{R}$ is the reward function. $\gamma$ is the discount rate. The objective of an agent in an MDP is to find an optimal policy ($\pi_{\theta}:\mathcal{S}\times \mathcal{A}\mapsto [0,1]$) which maximizes the expected cumulative rewards from any state $s\in \mathcal{S}$, i.e., $V^{*}(s)=\max_{\pi_{\theta}}\mathbb{E}_{\pi_{\theta}}\{\sum_{k=0}^{\infty}\gamma^{k}r_{t+k}|s_{t}=s\}$, or maximizes equivalently the expected cumulative rewards from any state-action pair $s\in\mathcal{S},a\in\mathcal{A}$, i.e., $Q^{*}(s,a)=\max_{\pi_{\theta}}\mathbb{E}_{\pi_{\theta}}\{\sum_{k=0}^{\infty}\gamma^{k}r_{t+k}|s_{t}=s,a_{t}=a\}$. Here $\mathbb{E}_{\pi_{\theta}}$ is the expectation under policy $\pi_{\theta}$, $t$ is the current timestep and $r_{t+k}$ is the immediate reward at a future timestep $t+k$.

We model the recommendation procedure as a sequential decision making problem, in which the recommender (i.e., agent) interacts with users (i.e., environment) to suggest a list of items sequentially over the timesteps, by maximizing the cumulative rewards of the whole recommendation procedure. More specifically, the recommendation procedure is modeled by an MDP, as follows.

\begin{itemize}
\item \textbf{States $\mathcal{S}$.} A state $s$ is the representation of user's positive interaction history with recommender, as well as her demographic information (if it exists in the datasets).
\item \textbf{Actions $\mathcal{A}$.} An action $a$ is a continuous parameter vector denoted as $a \in {\mathbb{R}^{1 \times k}}$. Each item $i_t \in {\mathbb{R}^{1 \times k}}$ \footnote{$i_t$ is the embedding of item $i$, which can be generated by MF or VAE.} has a ranking score, which is defined as the inner product of the action and the item embedding, i.e., ${i_t}{a^\top}$. Then the top ranked ones will be recommended.
\item \textbf{Transitions $\mathcal{P}$.} The state is modeled as the representation of user's positive interaction history. Hence, once the user's feedback is collected, the state transition is determined. 
\item \textbf{Reward $\mathcal{R}$.} Given the recommendation based on the action $a$ and the user state $s$, the user will provide her feedback, i.e., click, not click, or rating, etc. The recommender receives immediate reward $R(s,a)$ according to the user's feedback. 
\item \textbf{Discount rate $\gamma$.} $\gamma\in[0,1]$ is a factor measuring the present value of long-term rewards. In the case of $\gamma=0$, the recommender considers only immediate rewards but long-term rewards are ignored. On the other hand, when $\gamma=1$, the recommender treats immediate rewards and long-term rewards as equally important. 
\end{itemize}

\begin{figure}[ht]
\centering
\includegraphics[width=0.45\textwidth]{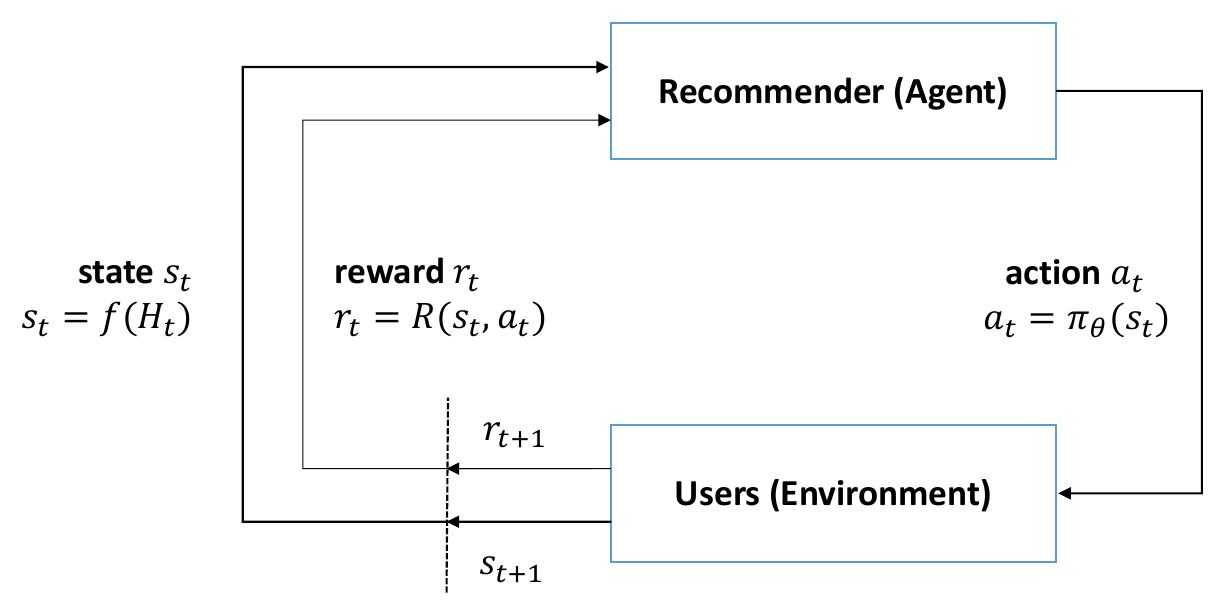}
\caption{Recommender-User interactions in MDP}\label{fig:rl}
\end{figure}

Figure~\ref{fig:rl} illustrates the recommender-user interactions in MDP formulation. Considering the current user state and immediate reward to the previous action, the recommender takes an action. Note that in our model, an action corresponds to neither recommending an item nor recommending a list of items. Instead, an action is a continuous parameter vector. Taking such an action, the parameter vector is used to determine the ranking scores of all the candidate items, by performing inner product with item embeddings. All the candidate items are ranked according to the computed scores and Top-N items are recommended to the user. Taking the recommendation from the recommender, the user provides her feedback to the recommender and the user state is updated accordingly. The recommender receives rewards according to the user's feedback. Without loss of generalization, a recommendation procedure is a $T$ timestep\footnote{If a recommendation episode terminates in less than T timesteps, then the length of the episode is the actual value.} trajectory as $(s_0,a_0,r_0, s_1,a_1,r_1, ..., s_{T-1},a_{T-1},r_{T-1}, s_{T})$.

\section{The Proposed DRR Framework}



As aforementioned in Section 1, conventional recommendation techniques suffer from either a lack of sequential modeling or ignoring the long-term rewards, or both. To address the drawbacks, we propose a deep reinforcement learning based recommendation framework (DRR) based on the Actor-Critic learning scheme. Also, different from some recent RL studies, we carefully and explicitly build a state representation module to model the interactions between the users and items. Next, we will first elaborate the Actor network, Critic network and the state representation module respectively, which are essentially the three key ingredients in our framework; then the training and evaluation procedures will be presented to show how to learn and use the DRR framework.

\begin{figure}[ht]
\centering
\includegraphics[width=0.48\textwidth]{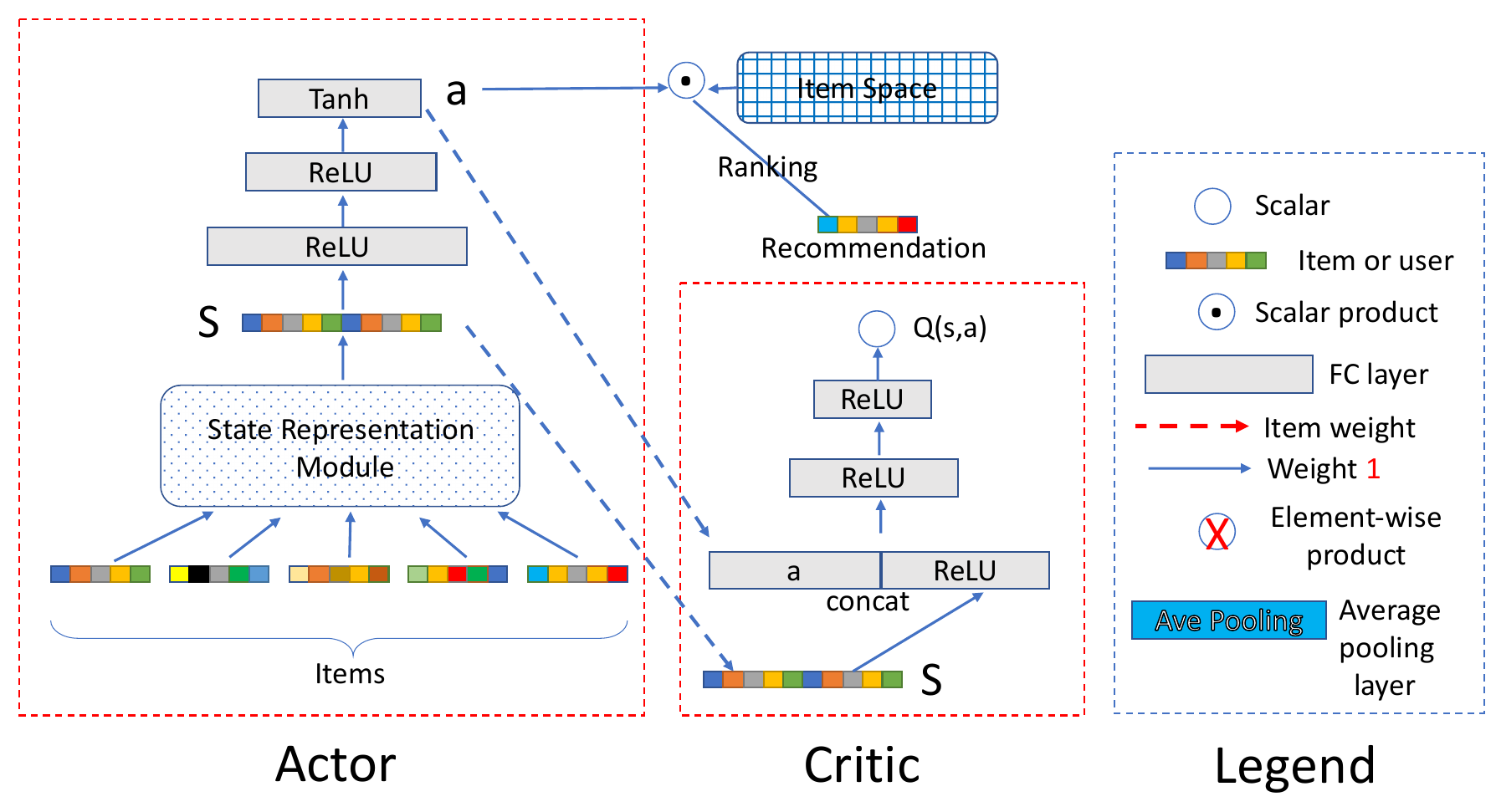}
\caption{DRR Framework}\label{fig:DRR}
\end{figure}

\subsection{Three Key Ingredients in DRR}

\subsubsection{The Actor network}

The Actor network, also called the policy network, is depicted on the left part of Figure~\ref{fig:DRR}. For a given user, the network accounts for generating an action $a$ based on her state $s$. Let us explain the network from the input to the output part. In DRR, the user state, denoted by the embeddings of her $n$ latest positively interacted items, is regarded as the input. Then the embeddings are fed into a state representation module (which will be introduced in details later) to produce a summarized representation $s$ for the user. For instance, at timestep $t$, the state can be defined in Eq. (\ref{eq:s}):


\begin{equation}\label{eq:s}
{s_t} = f({H_t})
\end{equation}

\noindent where $f( \cdot )$ stands for the state representation module, ${H_t} = \{ {i_1},...,{i_n}\}$ denotes the embeddings of the latest positive interaction history, and $i_t \in {\mathbb{R}^{1 \times k}}$ is a $k$-dimensional vector. When the recommender agent recommends an item $i_t$, if the user provides positive feedback, then in the next timestep, the state is updated to ${s_{t+1}} = f({H_{t+1}})$, where $H_{t+1} = \{ {i_2},...,{i_n},{i_t} \}$; otherwise, $H_{t+1} = H_t$. The reasons to define the state in such a manner are two folds: (i) a superior recommender system should cater to the users' taste, i.e., what items the users like; (ii) the latest records represent the users' recent interests more precisely. 

Finally, by two ReLU layers and one Tanh layer, the state representation $s$ is transformed into an action $a = {\pi _\theta }({s})$ as the output of the Actor network. Particularly, the action $a$ is defined as a ranking function represented by a continuous parameter vector $a \in {\mathbb{R}^{1 \times k}}$. By using the action, the ranking score of the item $i_t$ is defined as:

\begin{equation}\label{eq:score}
{score_t} = {i_t}{a^\top}
\end{equation}

\noindent Then, the top ranked item (w.r.t. the ranking scores) is recommended to the user. Note that, the widely used $\varepsilon$-greedy exploration technique is adopted here.


\subsubsection{The Critic network}
The Critic part in DRR, shown as the middle part of Figure~\ref{fig:DRR}, is a Deep Q-Network~\cite{davidHuman}, which leverages a deep neural network parameterized as ${Q_\omega }(s,a)$ to approximate the true state-action value function ${Q^\pi }(s,a)$, namely, the Q-value function. The Q-value function reflects the merits of the action policy generated by the Actor network. Specifically, the input of the Critic network is the user state $s$ generated by the user state representation module and the action $a$ generated by the policy network, and the output is the Q-value, which is a scalar. According to the Q-value, the parameters of the Actor network are updated in the direction of improving the performance of action $a$, i.e., boosting ${Q_\omega }(s,a)$. Based on the deterministic policy gradient theorem~\cite{dpg}, we can update the Actor by the sampled policy gradient shown in Eq.(\ref{eq: actor gradient in DRR}):

\begin{equation}\label{eq: actor gradient in DRR}
{\nabla _\theta }J(\pi _ {\theta})  \approx \frac{1}{N}\sum\nolimits_t {{\nabla _a}{Q_\omega }(s,a){|_{s = {s_t},{a = \pi _\theta }({s_t})}}{\nabla _\theta }{\pi _\theta }(s)} {|_{s = {s_t}}}
\end{equation}

\noindent where $J(\pi _ {\theta})$ is the expectation of all possible Q-values that follow the policy $\pi _{\theta}$. Here the mini-batch strategy is utilized and $N$ denotes the batch size. Moreover, the Critic network is updated accordingly by the temporal-difference learning approach~\cite{sutton1998reinforcement}, i.e., minimizing the mean squared error shown in Eq.(\ref{eq: mse in critic}):

\begin{equation}\label{eq: mse in critic}
L = \frac{1}{N}\sum\nolimits_i {{{({y_i} - {Q_\omega }({s_i},{a_i}))}^2}} 
\end{equation}

\noindent where ${y_i} = {r_i} + \gamma {Q_{\omega'}}({s_{i + 1}},{\pi _{\theta '}}({s_{i + 1}}))$. The target network~\cite{DDPG} technique is also adopted in DRR framework, where $\omega'$ and $\theta'$ is the parameters of the target Critic and Actor network. 

\subsubsection{The State Representation Module}
As noted above, the state representation module plays an important role in both the Actor network and Critic network. Hence, it is very crucial to design a good structure to model the state. In~\cite{pnn,deepfm}, it has been shown that modeling the feature interactions explicitly can boost the performance of a recommendation system. Inspired by the studies, we propose to design the state representation module by explicitly modeling the interactions between the users and items. Specifically, we develop three structures, which will be elaborated next.


\begin{figure}[ht]
\centering
\includegraphics[width=0.48\textwidth]{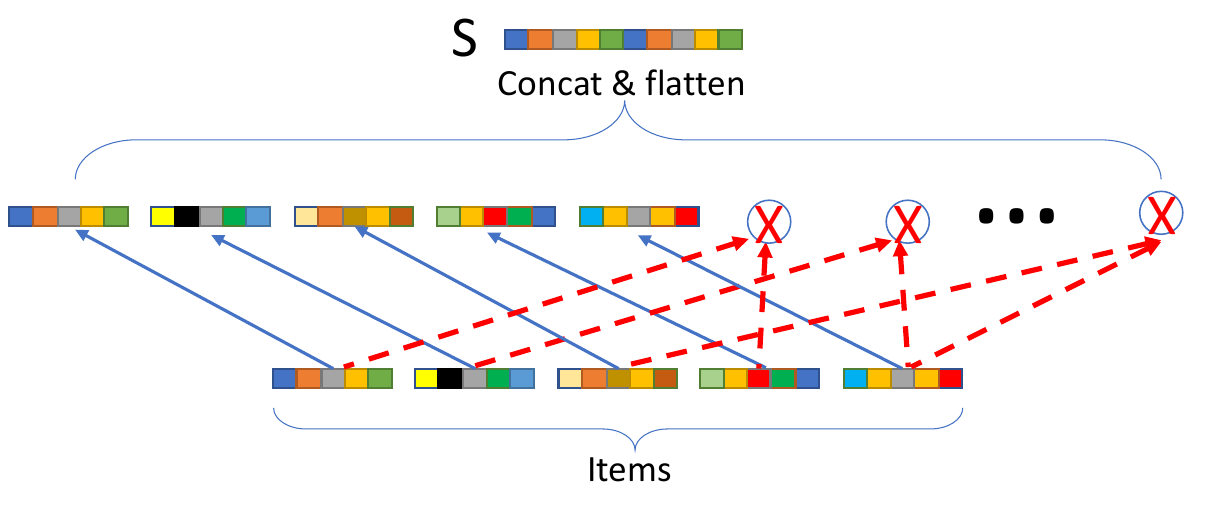}
\caption{DRR-p Structure}\label{fig:DRR-p}
\end{figure}

\begin{itemize}

\item \textbf{DRR-p}. Inspired by~\cite{pnn,deepfm}, we propose a product based neural network for the state representation module, which is depicted in Figure \ref{fig:DRR-p}\footnote{The legend in Figure \ref{fig:DRR-p}, \ref{fig:DRR-u} and \ref{fig:DRR-ave} is the same to Figure \ref{fig:DRR}}. The structure is named as DRR-p, which utilizes a product operator to capture the pairwise local dependency between items. We can see that the structure clones the representations of the $n$ items from ${H} = \{ {i_1},...,{i_n}\}$. In addition, it computes the pairwise interactions between the $n$ items, by using the element-wise product operator. As a result, $n(n-1)/2$ new features vectors are yielded, which will be concatenated with the cloned vectors as the state representation. We note that in the element-wise product part, a weight is also learned for each item to show its importance. Hence, in DRR-p the state representation module can be formally stated as follows:



\begin{equation}\label{eq: s_p}
s = [ H,\{ {p_{a,b}}|a,b = 1,...,n\}] 
\end{equation}

\begin{equation}\label{eq: p}
{p_{a,b}} = {w_a}{i_a} \otimes {w_b}{i_b}
\end{equation}

\noindent where $\otimes$ denotes the element-wise product, $w_a$ is a scalar indicating the importance of item $i_a$, and $p_{a,b}$ is a $k$-dimensional vector which models the interactions between item $i_a$ and $i_b$. The dimensionality of $s$ is $k(n + n(n-1)/2)$. 

\item \textbf{DRR-u}. Though DRR-p can model the pairwise local dependency between items, the user-item interactions are neglected. To remedy this, we design another structure in Figure \ref{fig:DRR-u}, which is referred as DRR-u. In DRR-u, we can see that the user embedding is also incorporated. In addition to the local dependency between items, the pairwise interactions of user-item are also taken into account. Formally, the state representation module can be expressed as:


\begin{equation}\label{eq: s_u}
s = [ \{u \otimes {w_a}{i_a}|a = 1,...,n\},\{ {p_{a,b}}|a,b = 1,...,n\} ] 
\end{equation}

\noindent The dimensionality of $s$ is also $k(n + n(n-1)/2)$.

\begin{figure}[ht]
\centering
\includegraphics[width=0.48\textwidth]{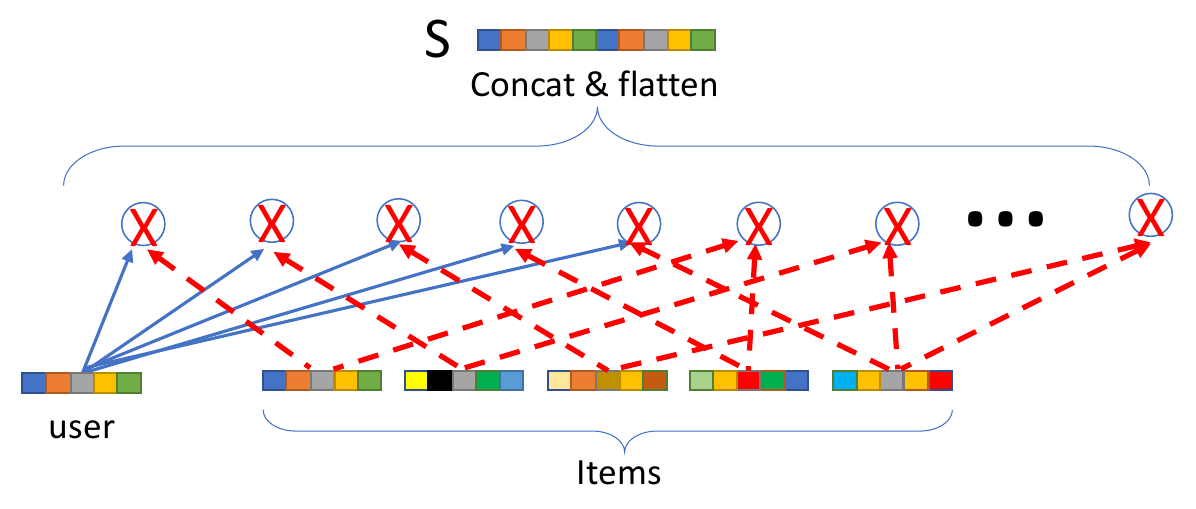}
\caption{DRR-u Structure}\label{fig:DRR-u}
\end{figure}

\item \textbf{DRR-ave}. In DRR-p and DRR-u structures, the interactions between users and items can be exploited and modeled. For the two structures, it is not difficult to find that the positions of items in $H$ matters, e.g., the state representations of $H_1=\{i_a, i_b, i_c\}$ and $H_2=\{i_c, i_b, i_a\}$ are different. When $H$ is large, we expect the positions of items really matter, because $H$ denotes a long-term sequence; whereas memorizing the positions of items may lead to overfitting if the sequence $H$ is a short-term one. Hence, we design another structure by eliminating the position effects, which is depicted in Figure \ref{fig:DRR-ave}. As an average pooling layer is adopted, we call the structure DRR-ave. We can see from Figure \ref{fig:DRR-ave} that the embeddings of items in $H$ are first transformed by a weighted average pooling layer. Then, the resulting vector is leveraged to model the interactions with the input user. Finally, the embedding of the user, the interaction vector, and the average pooling result of items are concatenate into a vector to denote the state representation. Formally, the DRR-ave structure can be expressed as:


\begin{equation}\label{eq: s_ave}
s = [ u,u \otimes \{g({i_a})|a = 1,...,n\},\{g({i_a})|a = 1,...,n\}]
\end{equation}

\begin{equation}\label{eq: g_H}
g({i_a}) = ave({w_a}{i_a})|a = 1,...,n
\end{equation}

\noindent Here $g(\cdot)$ indicates the weighted average pooling layer. The dimensionality of $s$ in DRR-ave is $3k$.

\begin{figure}[ht]
\centering
\includegraphics[width=0.4\textwidth]{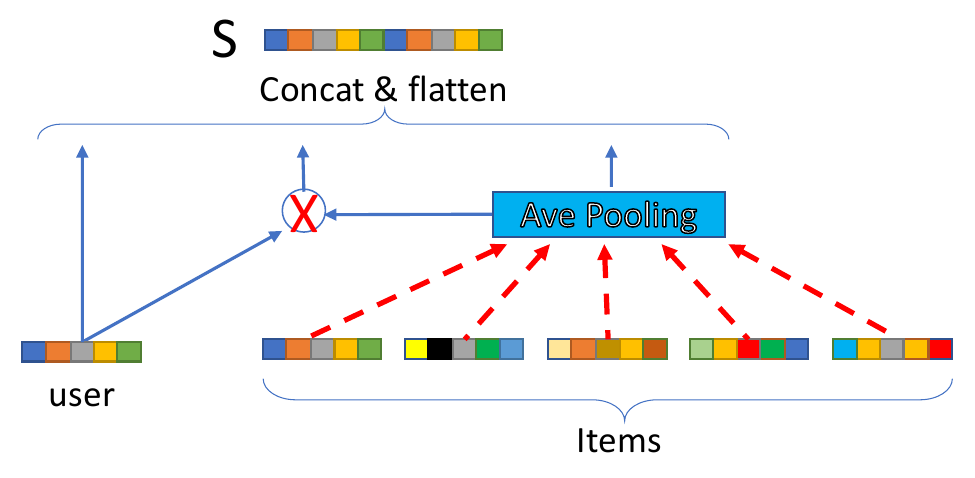}
\caption{DRR-ave Structure}\label{fig:DRR-ave}
\end{figure}

\end{itemize}

\subsection{Training Procedure of the DRR Framework}
Next, we introduce how to train the DRR framework. We first present the overall idea and then discuss the detailed algorithm. As aforementioned, DRR utilizes the users' interaction history with the recommender agent as training data. During the procedure, the recommender takes an action $a_t$ following the current recommendation policy ${\pi _\theta }({s_t})$ after observing the user (environment) state $s_t$, then it obtains the feedback (reward) $r_t$ from the user, and the user state is updated to $s_{t+1}$. According to the feedback, the recommender updates its recommendation policy. In this work, we utilize deep deterministic policy gradient (DDPG)~\cite{DDPG} algorithm to train the proposed DRR framework, as detailed in Algorithm 1. 

Specifically, in timestep $t$, the training procedure mainly includes two phases, i.e., transition generation (lines 7-12) and model updating (lines 13-17). For the first stage, the recommender observes the current state $s_t$ that is calculated by the proposed state representation module, then generates an action ${a_t} = {\pi _\theta }({s_t})$ according to the current policy $\pi _\theta$ with $\varepsilon$-greedy exploration, and recommends an item $i_t$ according to the action $a_t$ by Eq. (\ref{eq:score}) (lines 8-9). Subsequently, the reward $r_t$ can be calculated based on the feedback of the user to the recommended item $i_t$, and the user state is updated (lines 10-11). Finally, the recommender agent stores the transition $({s_t},{a_t},{r_t},{s_{t + 1}})$ into the replay buffer $D$ (line 12). 


In the second stage, the model updating, the recommender samples a minibatch of $N$ transitions with widely used prioritized experience replay~\cite{prioritized} sampling technique (line 13), which is essentially an importance sampling strategy. Then, the recommender updates the parameters of the Actor network and Critic network according to Eq. (\ref{eq: actor gradient in DRR}) and Eq. (\ref{eq: mse in critic}) respectively (line 14-16). Finally, the recommender updates the target networks' parameters with the soft replace strategy.

\begin{algorithm}
{\caption{Training Algorithm of DRR Framework}}
\SetKwInOut{Input}{input}\SetKwInOut{Output}{output}
    \Input{Actor learning rate $\eta_a$, Critic learning rate $\eta_c$, discount factor $\gamma$, batch size $N$, state window size $n$ and reward function $R$}
\nl    Randomly initialize the Actor $\pi _\theta$ and the Critic $Q_\omega$ with parameters $\theta$ and $\omega$\\
\nl    Initialize the target network $\pi'$ and $Q'$ with weights $\theta ' \leftarrow \theta$ and $\omega ' \leftarrow \omega$ \\
\nl    Initialize replay buffer $D$\\
\nl    \For {session = 1, M}{
\nl        Observe the initial state $s_0$ according to the offline log\\ 
\nl        \For{ t = 1, T}{
\nl     		Observe current state ${s_t} = f({H_t})$, where ${H_t} = \{ {i_1},...,{i_n}\}$\\
\nl        		Find action ${a_t} = {\pi _\theta }({s_t})$ according to the current policy with $\varepsilon$-greedy exploration\\
\nl 			Recommended item $i_t$ according to action $a_t$ by Eq. (\ref{eq:score})\\
\nl        		Calculate reward $r_t = R(s_t, a_t)$ based on the feedback of the user\\
\nl  			Observe new state ${s_{t+1}} = f({H_{t+1}})$, where $H_{t+1} = \{ {i_2},...,{i_n},{i_t} \}$ if $r_t$ is positive, otherwise, $H_{t+1} = H_t$\\
\nl 			Store transition $({s_t},{a_t},{r_t},{s_{t + 1}})$ in $D$\\
\nl 			Sample a minibatch of $N$ transitions $({s_i},{a_i},{r_i},{s_{i + 1}})$ in $D$ with prioritized experience replay sampling technique\\
\nl 			Set ${y_i} = {r_i} + \gamma {Q_{\omega'}}({s_{i + 1}},{\pi _{\theta '}}({s_{i + 1}}))$\\
\nl 			Update the Critic network by minimizing the loss: 
$L = \frac{1}{N}\sum\nolimits_i {{{({y_i} - {Q_\omega }({s_i},{a_i}))}^2}} $\\
\nl 			Update the Actor network using the sampled policy gradient: \\
${\nabla _\theta }J(\pi _ {\theta})  \approx \frac{1}{N}\sum\nolimits_t {{\nabla _a}{Q_\omega }(s,a){|_{s = {s_t},{a = \pi _\theta }({s_t})}}{\nabla _\theta }{\pi _\theta }(s)} {|_{s = {s_t}}}$\\
\nl 			Update the target networks:\\
$\theta ' \leftarrow \tau \theta  + (1 - \tau )\theta '$\\
$\omega ' \leftarrow \tau \omega  + (1 - \tau )\omega '$\\
			}
        }
\nl    \Return $\theta$ and $\omega$
\end{algorithm}

\subsection{Evaluation}

In this subsection, we discuss how to evaluate the models with a environment simulator. The most straightforward way to evaluate the RL based models is to conduct online experiments on recommender systems where the recommender directly interacts with users. However, the underlying commercial risk and the costly deployment on the platform make it impractical. Therefore, throughout the testing phase, we conduct the evaluation of the proposed models on public offline datasets and propose two ways to evaluate the models, which are the offline evaluation and the online evaluation. 

\subsubsection{Offline evaluation}
Intuitively, the offline evaluation of the trained models is to test the recommendation performance with the learned policy, which is described in Algorithm 2. Specifically, for a given session $\mathcal{S}_j$, the recommender only recommends the items that appear in this session, denoted as $\mathcal{I}(\mathcal{S}_j)$, rather than the ones in the whole item space. The reason is that we only have the ground truth feedback for the items in the session in the recoreded offline log. For each timestep, the recommender agent takes an action $a_t$ according to the learned policy $\pi _\theta$, and recommends an item $i_t \in \mathcal{I}(\mathcal{S}_j)$ based on the action $a_t$ by Eq. (\ref{eq:score}) (lines 4-5). After that, the recommender observes the reward $r_t = R(s_t, a_t)$ according to the feedback of the recommended item $i_t$ by Eq. (\ref{eq:reward function}) (lines 5-6). Then the user state is updated to $s_{t+1}$ and the recommended item $i_t$ is removed from the candidate set $\mathcal{I}(\mathcal{S}_j)$ (lines 7-8). The offline evaluation procedure can be treated as a rerank procedure of the candidate set by iteratively selecting an item w.r.t. the action generated by the Actor network in DRR framework. Moreover, the model parameters are not updated in the offline evaluation.


\begin{algorithm}
{\caption{Offline Evaluation Algorithm of DRR Framework}}
\SetKwInOut{Input}{input}\SetKwInOut{Output}{output}
    \Input{state window size $n$ and reward function $R$}
\nl Observe the initial state $s_0$ and item set $\mathcal{I}$ according to the offline log\\ 
\nl \For{ t = 1, T}{
\nl     Observe current state ${s_t} = \{ {i_1},...,{i_n}\}$\\
\nl     Execute action ${a_t} = {\pi _\theta }({s_t})$ according to the current policy \\
\nl 	Observe the recommended item $i_t$ according to action $a_t$ by Eq. (\ref{eq:score})\\
\nl     Get reward $r_t = R(s_t, a_t)$ from the feedback located in the users' log by Eq. (\ref{eq:reward function}) \\
\nl 	Update to a new state ${s_{t+1}} = f({H_{t+1}})$, where $H_{t+1} = \{ {i_2},...,{i_n},{i_t} \}$ if $r_t$ is positive, otherwise, $H_{t+1} = H_t$\\
\nl 	remove $i_t$ from $\mathcal{I}$\\
	}
\end{algorithm}

\subsubsection{Online evaluation with environment simulator} 
As aforementioned that it is risky and costly to directly deploy the RL based models on recommender systems. Therefore, we conduct online evaluation with an environment simulator. In this paper, we pretrain a PMF~\cite{PMF} model as the environment simulator, i.e., to predict an item's feedback that the user never rates before. The online evaluation procedure follows Algorithm 1, i.e., the parameters continuously update during the online evaluation stage. Its major difference from Algorithm 1 is that the feedback of a recommended item is observed by the environment simulator. Moreover, before each recommendation session starting in the simulated online evaluation, we reset the parameters back to $\theta$ and $\omega$ which is the policy learned in the training stage for a fair comparison. 

\section{Experiment}

\subsection{Datasets and Evaluation Metrics}

We adopt the following publicly available datasets from the real world to conduct the experiments:

\begin{itemize}

\item \textbf{MovieLens (100k)\footnote{https://grouplens.org/datasets/movielens/100k/}}. A benchmark dataset comprises of 0.1 million ratings from users to the recommended movies on MovieLens website.

\item \textbf{Yahoo! Music (R3)\footnote{https://webscope.sandbox.yahoo.com/}}. This dataset contains over 0.36 million ratings of songs collected from two different sources. The first source consists of ratings provided by users during normal interactions with Yahoo! Music services. The second source consists of ratings of randomly selected songs collected during an online survey by Yahoo! Research. We normalize the ratings to discrete values from 1 to 5.

\item \textbf{MovieLens (1M)\footnote{https://grouplens.org/datasets/movielens/1m/}}. A benchmark dataset includes of 1 million ratings from the MovieLens website.

\item \textbf{Jester (2)\footnote{http://eigentaste.berkeley.edu/dataset/}}. This dataset contains over 1.7 million real-value ratings (-10.0 to +10.0) over jokes in an online joke recommender system. 
\end{itemize} 

Note that except for Jester, the ratings in the other datasets are discrete values from 1 to 5, and the statistic information of the datasets is given in Table \ref{tab: datasets}. The MovieLens (100k) and MovieLens (1M) are abbreviated as ML (100k) and ML (1M) respectively.

\begin{table}[ht]
  \centering
  \caption{Statistic information of the datasets}
  \begin{tabular}{c|ccccccc}
    \toprule
    	               & \bf{ML (100k)} & \bf{Yahoo! Music}   & \bf{ML (1M)} & \bf{Jester} \\
    \midrule
    \# user            & 943            & 15,400              & 6,040        & 63,978     \\
    \# item            & 1,682          & 1,000               & 3,952        & 150     \\
    \# ratings         & 100,000        & 365,740             & 1,000,209    & 1,761,439   \\
    \bottomrule
  \end{tabular}
  \label{tab: datasets}
\end{table}

We conduct both offline and simulated online evaluation on these four datasets. For the offline evaluation, we utilize \textbf{Precision@k} and \textbf{NDCG@k} as the metrics to measure the performance of the proposed models. For the simulated online evaluation, we leverage the total accumulated rewards as the metric.

\subsection{Compared Methods}

We compare the proposed methods with some representative baseline methods. For the offline evaluation, we compare to conventional methods including Popularity, PMF~\cite{PMF} and SVD++~\cite{SVD++}, and a RL based method DRR-n. Moreover, the online evaluation baselines contain the state-of-the-art multi-armed bandits methods LinUCB~\cite{LinUCB} and HLinUCB~\cite{HLinUCB} and the DRR-n as well. 

\begin{itemize}

\item \textbf{Popularity} recommends the most popular item, i.e., the item with the highest average rating or the items with largest number of positive ratings\footnote{To get a better result of popularity based recommendation, we both test the two strategies, and choose the best one to report.} from current available items to the users at each timestep. 

\item \textbf{PMF} makes a matrix decomposition as SVD, while it only takes into account the non zero elements.

\item \textbf{SVD++} mixes strengths of the latent model as well as the neighborhood model.

\item \textbf{LinUCB} selects an arm (item) according to the estimated upper confidence bound of the potential reward. 

\item \textbf{HLinUCB} further learns hidden features for each arm to model the potential reward.

\item \textbf{DRR-n} simply utilizes the concatenation of the item embeddings to represent user state, which is widely used in previous studies. Although it is under the DRR framework, we treat this method as a baseline to assess the effectiveness of our proposed state representation module.

\end{itemize}







\subsection{Experimental Settings}
For each dataset, we choose 80\% of the interactions in each user session as the training set, and leave the rest as the testing set. Moreover, for MovieLens (100k), Yahoo! Music and MovieLens (1M), the positive ratings are $4$ and $5$, while for Jester, the positive ones are those higher than $0$. The number of latest positively rated items $n$, which is empirically set to $5$. We perform PMF to pretrain the $100$-dimensional embeddings of the users and items. Moreover, in each episode, we do not recommend repeated items, i.e., we remove the ones already recommended from the candidate set. The discount rate $\gamma$ is 0.9. We utilize Adam optimizer for all the RL based methods with $L_2$-norm regularization to prevent overfitting. As for the reward function, we empirically normalize the ratings into range [-1 ,1] and utilize the normalized ones as the feedback of the corresponding recommendations. For instance, in timestep $t$, the recommender agent recommends an item $j$ to user $i$, (denoted as action $a$ in state $s$), and the rating $rate_{i,j}$ comes from the interaction logs if user $i$ actually rates item $j$, or from a predicted value by the simulator otherwise. Therefore, the reward function can be defined as follows:

\begin{equation}\label{eq:reward function}
\begin{aligned}
&R(s,a) = \frac{1}{2}({rate}_{i,j} - 3) \\
&R(s,a) = {rate}_{i,j}/10
\end{aligned}
\end{equation}

\noindent where the first setting is for MovieLens (100k), Yahoo! Music and MovieLens (1M), and the second one is for Jester. All the baseline methods are carefully tuned for a fair comparison. We model the recommendation procedure as an interaction episode with length $T$, and the hyper-parameter $T$ is tuned for different datasets (detailed in Section V.E).

\subsection{Results and Analysis}

\subsubsection{Offline Evaluation Results and Analysis}
The offline evaluation results are summarized from Table \ref{tab: results ml100k} to Table \ref{tab: results jester} respectively, where the best results are marked in bold type. In the offline evaluation, we compare the proposed methods to some representative offline learning methods. The results suggest that the proposed methods under the DRR framework outperform the baselines on most of datasets, which demonstrates the effectiveness of our proposed methods.

Specifically, as aforementioned that, we propose three different network structure in the state representation module to model the explicit interactions of the users and items under the DRR framework, which are the DRR-p, DRR-u and DRR-ave. From the results in Table \ref{tab: results ml100k} to Table \ref{tab: results jester}, we find that the three methods all outperform the baselines in most cases. Moreover, DRR-n that simply concats the item embeddings to represent the state $s$, performs worse than the proposed DRR-p, DRR-u and DRR-ave. From the observations, we can conclude in two folds: (i) the proposed methods indeed have the capability of long-term scheduling and dynamic adaptation, which are ignored by conventional methods; (ii) the proposed state representation module well captures the dynamic interactions between the users and items, and the state should not be simply concatenate with fully connected layers as DRR-n does, which may result in information loss.


Compared with DRR-p, DRR-u and DRR-ave, we can see that DRR-ave outperforms DRR-u, and DRR-u is superior than DRR-p on the four datasets in most cases. The reasons are as follows: 1) The DRR-u method has better performance than DRR-p, because DRR-u only captures the interactions of user's historical items, but also seizes the personalization information through the user-item interactions. 2) DRR-ave performs the best, because of two reasons: (i) DRR-ave method captures the personalization information through user-item interactions; (ii) as noted in Section IV, by using the average pooling, it eliminates the position effects in $H$.



\begin{table}[ht]
  \centering
  \caption{Performance comparison of all methods on \textbf{ML (100k)} dataset.}
  \begin{tabular}{c|cccc}
    \toprule
    Model	           & Precision@5  & Precision@10  & NDCG@5  & NDCG@10 \\
    \midrule
    Popularity         & 0.6933     & 0.6012      &  0.9104     & 0.9008  \\
    PMF            	   & 0.6988     & 0.6194      &  0.9095     & 0.8968  \\
    SVD++              & 0.7034     & 0.6255      &  0.9125     & 0.8991  \\
    \midrule
    DRR-n			   & 0.7185     & 0.6387      &  0.9147     & 0.9004  \\
    \midrule
    DRR-p			   & 0.7263     & 0.6448      &  0.9076     & 0.9015  \\
    DRR-u			   & 0.7417     & 0.6536      &  0.9183     & \bf{0.9062}  \\
    DRR-ave			   & \bf{0.7887}     & \bf{0.6935}      &  \bf{0.9255}     & 0.9046  \\
    \bottomrule
  \end{tabular}
  \label{tab: results ml100k}
\end{table}

\begin{table}[ht]
  \centering
  \caption{Performance comparison of all methods on \textbf{Yahoo! Music} dataset.}
  \begin{tabular}{c|cccc}
    \toprule
    Model	           & Precision@5   & Precision@10  & NDCG@5 & NDCG@10 \\
    \midrule
    Popularity         & 0.3826      & 0.3805     & 0.8870  & 0.8811     \\
    PMF            	   & 0.3835      & 0.3817     & 0.8837  & 0.8802     \\
    SVD++              & 0.3857      & 0.3821     & 0.8887  & 0.8813   \\
    \midrule
    DRR-n			   & 0.3844      & 0.3819     & 0.8876  & 0.8810  \\
    \midrule
    DRR-p			   & 0.3850      & 0.3822     & 0.8883  & 0.8815  \\
    DRR-u			   & 0.3864      & 0.3827     & 0.8889  & 0.8819\\
    DRR-ave			   & \bf{0.3917} & \bf{0.3839} & \bf{0.9004} & \bf{0.8949}\\
    \bottomrule
  \end{tabular}
  \label{tab: results yahoo}
\end{table}

\begin{table}[ht]
  \centering
  \caption{Performance comparison of all methods on \textbf{ML (1M)} dataset.}
  \begin{tabular}{c|cccc}
    \toprule
    Model	           & Precision@5  & Precision@10 & NDCG@5  & NDCG@10 \\
    \midrule
    Popularity         & 0.7141      & 0.6181       & 0.8906      & 0.8738  \\
    PMF            	   & 0.7072      & 0.6193       & 0.8901      & 0.8746  \\
    SVD++              & 0.7142      & 0.6258       & 0.9009      & 0.8776  \\
    \midrule
    DRR-n			   & 0.7151      & 0.6221       & 0.8902      & 0.8751  \\
    \midrule
    DRR-p			   & 0.7346      & 0.6366       & 0.8909      & 0.8753  \\
    DRR-u			   & 0.7375      & 0.6385       & 0.8912      & 0.8763  \\
    DRR-ave			   & \bf{0.7693} & \bf{0.6594}  & \bf{0.9112} & \bf{0.8980}\\
    \bottomrule
  \end{tabular}
  \label{tab: results 1m}
\end{table}

\begin{table}[ht]
  \centering
  \caption{Performance comparison of all methods on \textbf{Jester} dataset.}
  \begin{tabular}{c|cccc}
    \toprule
    Model	           & Precision@5   & Precision@10     & NDCG@5       & NDCG@10 \\
    \midrule
    Popularity         & 0.6167        & 0.6012           & 0.8932       & 0.8703   \\
    PMF            	   & 0.6171        & 0.6015           & 0.8740       & 0.8676   \\
    SVD++              & 0.6184        & 0.6027           & 0.8819       & 0.8614   \\
    \midrule
    DRR-n			   & 0.6178        & 0.6021           & 0.8915       & 0.8724   \\
    \midrule
    DRR-p			   & 0.6181        & 0.6029           & 0.8934       & 0.8753   \\
    DRR-u			   & 0.6217        & 0.6043           & 0.8974       & 0.8805   \\
    DRR-ave			   & \bf{0.6278}   & \bf{0.6076}      & \bf{0.9124}  & \bf{0.9079} \\
    \bottomrule
  \end{tabular}
  \label{tab: results jester}
\end{table}

\subsubsection{Simulated online evaluation results and analysis}
The results of the simulated online evaluation are summarized in Table \ref{tab: results online}, where the best results are marked in bold type. In the experiment, we only compare with the baseline methods that can perform online learning, which are LinUCB, HLinUCB and DRR-n. Again, we find that the proposed methods deliver higher rewards than all the baselines.

On the one hand, the fact suggests that the proposed RL-based methods model dynamic adaptation and long-term rewards better than the multi-armed bandits based methods LinUCB and HLinUCB. On the other hand, the observation indicates that the proposed state representation structures are superior to the naive full-connected network in DRR-n. Again, we observe that DRR-ave performs the best among all the three proposed interaction modeling structures.



\begin{table}[ht]
  \centering
  \caption{The rewards of all methods on the four datasets.}
  \begin{tabular}{c|cccc}
    \toprule
    Model	           & \bf{ML (100k)} & \bf{Yahoo! Music} & \bf{ML (1M)} & \bf{Jester} \\
    \midrule
    LinUCB             & 1,958           & 30,462.5           & 30,174        & 141,358.4       \\
    HLinUCB            & 1,475           & 32,725             & 32,785.5      & 147,105.5       \\
    \midrule
    DRR-n			   & 2,654.5         & 35,382.5           & 35,860        & 165,844.5    \\
    \midrule
    DRR-p			   & 2,832           & 37,328.5           & 36,653        & 177,414.2    \\
    DRR-u			   & 2,869           & 42,174.5           & 37,615        & 183,517.6    \\
    DRR-ave			   & \bf{3,251.5}    & \bf{49,095}        & \bf{40,588}   & \bf{194,860.7}    \\
    \bottomrule
  \end{tabular}
  \label{tab: results online}
\end{table}

\subsection{Parameter Study}
\begin{figure}[ht]
\centering
\begin{minipage}[b]{0.5\textwidth}\centering
\includegraphics[width=0.45\textwidth]{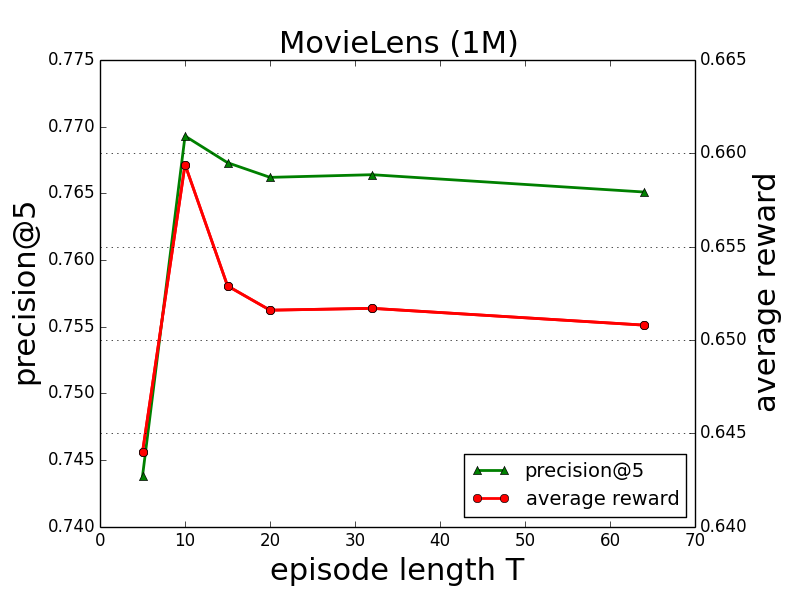}
\includegraphics[width=0.45\textwidth]{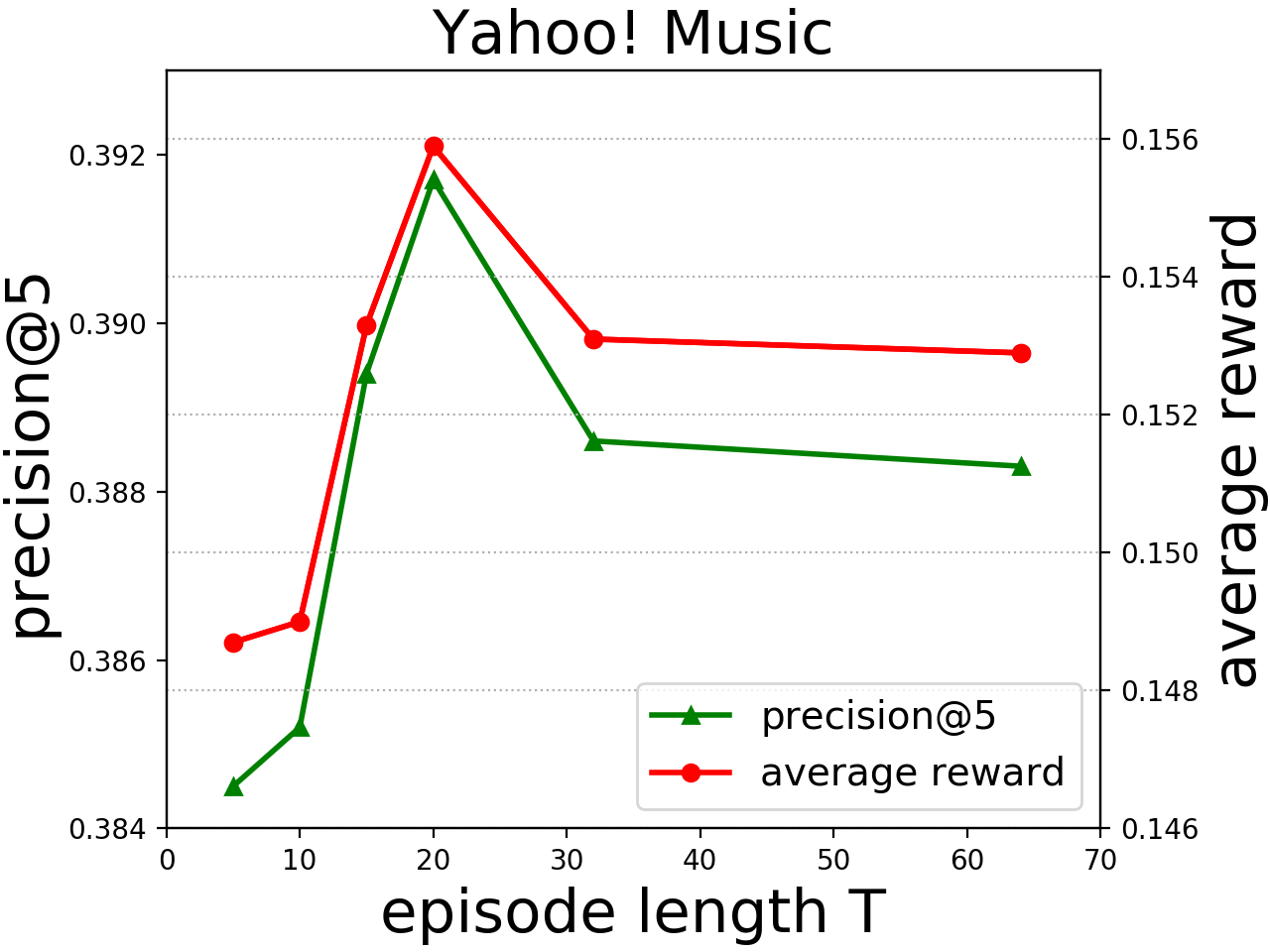}
\end{minipage}
\caption{Parameter study on episode length $T$ in MovieLens and Yahoo!Music datasets}\label{fig:episode_length}
\end{figure}

In this subsection, we investigate how the episode length $T$ affect the performance of proposed methods. Figure \ref{fig:episode_length} shows the results\footnote{Due to the space limit, We only present the performance of DRR-ave, while DRR-p and DRR-u have similar observations}. From the left part of Figure \ref{fig:episode_length}, we observe that the performance on MovieLens first increases and then decreases as the length of the episode is gradually increased, and the summit appears at $T=10$. A similar tend can be found for the Yahoo! Music from the right part of Figure \ref{fig:episode_length}, where the performance peaks at $T=20$. The reason may due to the trade-off between the exploitation and exploration. When the episode length is small, the user can not fully interact with the recommender agent, i.e., the exploration is insufficient. As we enlarge the episodes, the recommender agent can explore (interact with users) adequately, i.e., the recommender agent captures the user's preference, so that the performance improves. However, if the episodes are too large, the recommender focuses on exploiting locally, but the user preferred items is limited, therefore the performance declines as we do not recommend repeated items to user. Hence, we should nicely trade off the exploration and exploitation by setting a suitable value for $T$.



\subsection{Case Study}

In this subsection, we present an example to show the different recommendation manner between LinUCB and DRR-ave on MovieLens dataset. Specifically, we randomly pick up a user with ID 11, and conduct the recommendation procedure with LinUCB and DRR-ave respectively. To verify the reaction to the same recommendation scenario, we fix the first three recommended items and to see what will happen next. The results of recommended item and the reward are reported in Table \ref{tab: rec manner}.

From Table \ref{tab: rec manner}, we can see that LinUCB and DRR-ave react differently when given two consecutive negative recommendations (Eraser and First Knight). Specifically, LinUCB keeps exploring without considering to recommend a ``safe'' item to please the user. However, DRR-ave stops exploration and recommends a risk-free movie Dead Man Walking, which belongs to the same genre as Chasing Amy that has gained a positive feedback from the user at timestep 1. The observation demonstrates the superiority of the proposed DRR-ave against LinUCB.


\begin{table}[ht]
  \centering
  \caption{Different recommendation manner between LinUCB and DRR-ave on MovieLens. (The value in $(\cdot)$ denotes the corresponding reward.)}
  \begin{tabular}{c|cccc}
    \toprule
    timestep	           & LinUCB                  & DRR-ave    \\
    \midrule
    1                      & Chasing Amy (1)         & Chasing Amy (1)  \\
    2            	       & Eraser (-0.5)           & Eraser (-0.5)  \\
    3                      & First Knight (-1)       & First Knight (-1)   \\
    \midrule
    4			           & The Deer Hunter (-0.5)  & Dead Man Walking (1)   \\
    5			           & Event Horizon (-1)      & Braveheart (0.5)   \\
    6			           & The Net (0)             & The Usual Suspect (-0.5)   \\
    7			           & Striptease (-0.5)       & Psycho (0.5) \\
    \bottomrule
  \end{tabular}
  \label{tab: rec manner}
\end{table}

\section{Conclusion}
In this paper, we propose a deep reinforcement learning based framework DRR to perform the recommendation task. Unlike the conventional studies, DRR treats the recommendation as a sequential decision making process and adopts an ``Actor-Critic'' learning scheme, which can take both the immediate and long-term rewards into account. In DRR, a state representation module is incorporated and three instantiation structures are designed, which can explicitly model the interactions between users and items. Extensive experiments on four real-world datasets demonstrate the superiority of the proposed DRR method over state-of-the-art competitors.


\bibliographystyle{IEEEtran}
\bibliography{rlbib}

\end{document}